# Hematite $\alpha$-Fe$_2$O$_3$(0001) in top and side view: resolving long-standing controversies about its surface structure

*Jesús Redondo[1,2]\*, Jan Michalička[3], Giada Franceschi[4], Břetislav Šmid[1], Nishant Kumar[3], Ondřej Man[3], Matthias Blatnik[3], Dominik Wrana[1], Florian Kraushofer[4], Benjamin Mallada[5], Martin Švec[5], Gareth S. Parkinson[4], Martin Setvin[1], Michele Riva[4], Ulrike Diebold[4]\*, Jan Čechal[3]\**

AUTHOR ADDRESS:

1 Charles University, Faculty of Mathematics and Physics, Prague, Czech Republic

2 University of the Basque Country, Faculty of Chemistry, Donostia-San Sebastián, Spain

3 Central European Institute of Technology, Brno, Czech Republic

4 Institute for Applied Physics, Technical University Wien, Vienna, Austria

5 Institute of Physics, Czech Academy of Science, Prague, Czech Republic

AUTHOR INFORMATION

**Corresponding Author**

\*Jesús Redondo redondo@karlov.mff.cuni.cz.

\*Ulrike Diebold diebold@iap.tuwien.ac.at.

\*Jan Čechal cechal@fme.vutbr.cz

ABSTRACT.

Hematite $\alpha$-Fe$_2$O$_3$(0001) is the most-investigated iron oxide model system in photo and electrocatalytic research. The rich chemistry of Fe and O allows for many bulk and surface transformations, but their control is challenging. This has led to controversies regarding the structure of the topmost layers. This comprehensive study combines surface methods (nc-AFM, STM, LEED, and XPS) complemented by structural and chemical analysis of the near-surface bulk (HRTEM and EELS). The results show that a compact 2D layer constitutes the topmost surface of $\alpha$-Fe$_2$O$_3$(0001); it is locally corrugated due to the mismatch with the bulk. Assessing the influence of naturally-occurring impurities shows that these can force the formation of surface phases that are not stable on pure samples. Impurities can also cause the formation of ill-defined inclusions in the subsurface and modify the oxidation phase diagram of hematite. The results provide a significant step forward in determining the hematite surface structure that is crucial for accurately modeling catalytic reactions. Combining surface and cross-sectional imaging provided the full view that is essential for understanding the evolution of the near-surface region of oxide surfaces under oxidative conditions.

TOC.

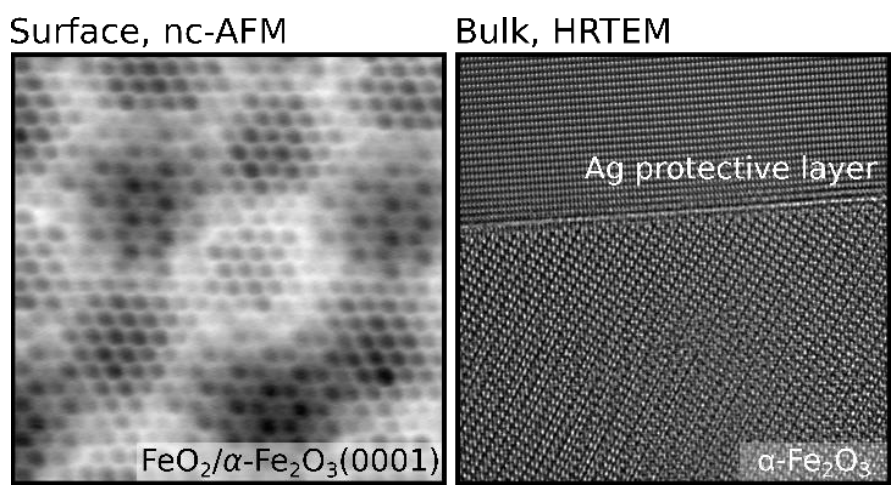

MAIN TEXT.

Iron oxides are extensively studied thanks to their plethora of electronic, magnetic, and catalytic properties conferred by the different crystal structures and stoichiometries of iron and oxygen in various compounds.[1] Hematite ($\alpha$-Fe$_2$O$_3$), the most abundant iron oxide on the Earth, has attracted much attention in green chemistry due to its large availability and low toxicity.[2] It is a promising photocathode material for the catalytic production of H$_2$ and CO$_2$-neutral solar-energy harvesting via photoelectrochemical water splitting (PWS).[3–5] Its bandgap (1.9–2.2 eV) permits electron excitations by visible light, but the actual efficiency of the PWS on hematite is well below the predicted (and industrially required) 15% solar-to-hydrogen conversion.[6] Sixty years of PWS research has led to the consensus that the reaction is promoted by electronic states formed at the interface between the oxide and the electrolyte.[7] However, the origin, amount, and role of these states are still a matter of controversy.[8,9] A comprehensive understanding of the surface atomic structure is crucial for the identification of active surface sites. Reaction-limiting factors such as light-depth penetration and charge-carrier trapping and recombination are inherently linked to the subsurface and bulk properties.[10,11] In recent years, attention has been paid to the characterization of bulk phenomena such as polaron formation and electron-hole recombination[12–14] and the effect of the bulk morphology on the final reaction yield.[15,16] Due to the importance of hematite in catalysis, the fundamental investigation of its redox behavior is the first step to designing improved industrial processes and chemical pathways.

The most investigated facet of hematite in catalysis is $\alpha$-Fe$_2$O$_3$(0001) (Fe$_2$O$_3$ for simplicity). It also is the most challenging of all iron oxide surfaces. The bulk and surface of natural Fe$_2$O$_3$ crystals undergo a wide range of phase transformations depending on the sample history, treatment, and, in the case of thin films, crystalline support.[17–19] Under low oxygen chemical potential, i.e., $\mu_o < -2.5$ eV, or a reducing agent, $Fe_2O_3$ single crystals form a thick layer of inverse-spinel $Fe_3O_4(111)$ with a bulk-terminated (1×1) surface structure.[20] Equilibrating (i.e., annealing under appropriate $O_2$ partial pressures) at chemical potentials > −2.5 eV initiates the stochiometric recovery towards $Fe_2O_3$. Different bulk phases have been identified in the −2.5 eV < $\mu_o$ < −1.8 eV range, namely magnetite ($Fe_3O_4$,) maghemite ($\gamma$-$Fe_2O_3$), and $\alpha$-$Fe_2O_3$.[21,22] The complexity of the bulk phases at this $\mu_o$ range is reflected at the surface: the identification of the structure of the $\alpha$-$Fe_2O_3(0001)$ surface has been a matter of a 40-year-long controversy. The two leading models are the 'biphase reconstruction', which consists of alternating domains with FeO and $Fe_2O_3$ stoichiometries[23], and the 'honeycomb model', an O–M–O– trilayer surface structure proposed as a possible polarity-compensation mechanism of corundum oxides.[24,25] Above $\mu_o = -1.8$ eV, the bulk fully recovers its $Fe_2O_3$ stoichiometry, and the surface is thought to terminate in a bulk-like (1×1) structure.[26] The $O_3$–, O–H–Fe-, and half-metal Fe terminations have been proposed for the (1×1) surface phase of $Fe_2O_3(0001)$.[27,28]

The coexistence of different bulk stoichiometries and surface phases on single crystals in ultra-high vacuum (UHV) hampers obtaining well-defined samples, presenting thus a "minefield"[1] in basic research of model hematite systems. Mixed bulk and surface iron oxide phases can be misidentified due to similarities in their spectral signatures, such as the Fe 2p $Fe^{3+}$ satellite structures of $Fe_2O_3$ and $\gamma$-$Fe_2O_3$ in X-ray photoemission spectroscopy (XPS)[29] or their pre-peak feature at the Fe 2p $L_3$ edge in X-ray absorption spectroscopy.[30] Thus, a combination of real- and reciprocal-space imaging is required to ensure the existence of a single surface phase.[24] Moreover, the presence of natural dopants (Na, Mg, K, Ca, Ti, V, Cr) in commercially available natural hematite crystals can shift the thermodynamic stability window of bulk hematite phases.[31] Surfaces make no exception: Foreign impurities can stabilize new surface phases. For example, K and Ti doping promotes (3×2) and (2×1) surface reconstructions on $Fe_2O_3(1\bar{1}02)$ for oxygen chemical potentials at which the (1×1) bulk reconstruction occurs in pristine hematite.[32,33] Sample impurities have also been linked to the (1×1) termination of $Fe_2O_3$[34] and α-$Al_2O_3$(0001).[35]

The rich chemistry of hematite requires analyzing the bulk and surface of iron oxide not as separated entities but as closely interrelated systems.[6] This work aims at solving the current challenges using α-$Fe_2O_3$(0001) as a model system for UHV investigations[36], and settle the controversy over the nature of the honeycomb/biphase reconstruction. The structural and chemical evolution of the bulk and surface of $Fe_2O_3$ single crystals was monitored at different redox conditions, also emphasizing the role of natural impurities. The surface of $Fe_2O_3$ was characterized by scanning tunneling microscopy (STM), non-contact atomic force microscopy (nc-AFM), low-energy electron diffraction (LEED), and XPS. Investigating the evolution of surface reconstructions on natural and synthetic $Fe_2O_3$ crystals allows disentangling the effect of sample impurities on their stability. High-resolution transmission electron microscopy (HRTEM) and electron energy loss spectroscopy (EELS) provide an atomically resolved view of the near-surface bulk phases of $Fe_2O_3$ single crystals. nc-AFM shows that the honeycomb/biphase $Fe_2O_3$ surface phase is formed by a compact 2D layer that is structurally perturbed locally. This local perturbation gives rise to the areas of distinct contrast identified as a 'biphase' in previous works.[23,37]

**Hematite bulk oxidation**

The near-surface bulk of $Fe_2O_3$ is significant for on-surface catalytic reactions as it can store defects, host electron–hole pair creation and their transport to the surface or determine the penetration depth of light. Model hematite systems used for UHV investigations lack information on the extent to which the bulk transforms when equilibrating at different oxygen chemical potentials. Mixed bulk phases and the interface between bulk hematite and surface phases cannot be probed by surface-sensitive techniques. To fully describe the surface, one needs information about the near-surface bulk structure.

Bulk phases of hematite thin films grown on polar oxides can be probed by HRTEM.[38] Here, HRTEM is used to access cross-sectional information on the sub-surface region of $Fe_2O_3$(0001) single crystals. Two limiting cases are presented: one sample featuring the $Fe_3O_4$(111) phase ($\mu_o < -2.5$ eV), and one fully $Fe_2O_3$ stoichiometrically recovered ($\mu_o > -1.8$ eV). As a technical note, the study of the sub-surface microstructure by TEM requires an electron-transparent specimen cut from the sample (a lamella). The fabrication of lamellas and the procedure to avoid surface contamination in the process is detailed in the methodology section.

The HRTEM images of the reduced sample (Figs. 1a, b) show the near-surface structure of the $Fe_3O_4$(111)-terminated $Fe_2O_3$. Two distinct regions are divided by a sharp interface. The first region is a surface layer that can be identified as a $Fe_3O_4$ lattice oriented to the zonal axis (i.e., parallel to the electron beam) [$\bar{1}12$], Fig. 1c, by evaluation of FFT and crystallographic standard.[39] Note that HRTEM cannot unambiguously distinguish the lattice of $Fe_3O_4$ from $\gamma$-$Fe_2O_3$ due to their similar inverse cubic spinel crystal structure and lattice parameter. The second region is a single-crystalline bulk with a dumbbell atomic pattern identified as the rhombohedral lattice of $Fe_2O_3$ hematite oriented to the zonal axis [$11\bar{2}0$], Fig. 1d, by evaluation of FFT and crystallographic standard.[40] Along the whole length of the lamella, the surface layer reveals a defect-free single crystal structure without fragmentation to misoriented sub-grains. The thickness of the surface layer is about 20–25 nm; however. This thickness generally depends on the degree of reduction of the sample, i.e., by the ion beam energy, total sputtering dose, and the duration of UHV-annealing treatments performed prior to cutting the lamellae. The HRTEM analysis of the fully reoxidized sample given in Figs. 1e and 1f reveals that $Fe_2O_3$, again identified by FFT evaluation and crystallographic standard, is present up to the sample surface, forming an atomically sharp interface with the Ag protective layer. This indicates that a full oxidation of the sample can be achieved at $\mu_o > -1.8$ eV even in a UHV chamber.

EELS mapping performed in scanning-TEM mode (STEM-EELS) provides spatially resolved information about the chemical composition, oxidation, and valence states of the investigated iron oxides.[41] Fig. 1g shows the EELS atomic concentration maps associated with Fe-L and O-K edges, obtained from the quantification of a spectrum image at the Ag-buried interface. The relative Fe and O concentration as a function of depth (along the arrow in Fig. 1g) is shown in Fig. 1h. The deeper bulk part shows a 40:60 Fe:O ratio, while a 43:57 ratio is measured in the upper surface layer. These values correspond well to the nominal stoichiometries expected for $Fe_2O_3$ (40:60) and $Fe_3O_4$ (42.9:57.1), respectively. Fig. 1i shows monochromated STEM-EELS spectra of the Fe-$L_{3,2}$ fine-edge structure measured in both bulk (cyan curve) and reduced (blue curve) regions of the same sample. The bulk spectrum shows a strong pre-peak $L_3$ component and two $L_2$ main contributions, whereas the upper layer shows no $L_3$ pre-peak and three $L_2$ contributions. These spectra are consistent with those expected for hematite and magnetite (not maghemite), respectively.[42] The combination of TEM and EELS structural and chemical analyses reveal the formation of a surface magnetite layer with a well-defined interface on the hematite bulk structure under the conditions of controlled UHV experiments.

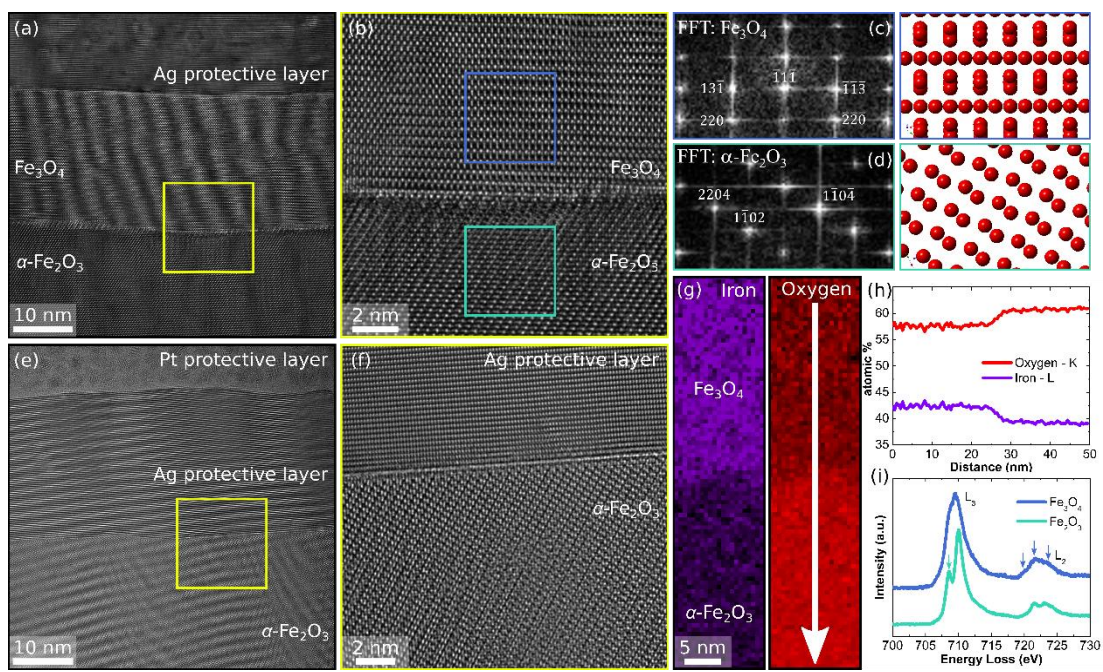

**Figure 1**. Reduced and oxidized hematite in side view. (a) and (b) HRTEM images of the partially reduced hematite showing an Fe3O4 magnetite film on Fe2O3 hematite substrate. The magnetite and hematite are oriented to the zonal axis $[\bar{1}12]$ and $[11\bar{2}0]$, respectively, as revealed by FFT analysis and structural modelling in panels (c) and (d). (e) and (f) HRTEM images of the fully oxidized sample revealing a single crystal structure identified by FFT as Fe2O3 hematite oriented to the $[11\bar{2}0]$ zonal axis. The dumbbell atomic structure characteristic of α-Fe2O3(0001) appears undisturbed all the way up the interface with the Ag protective layer. (g) STEM-EELS atomic concentration maps associated with Fe-L2,3 and O-K edges of the magnetite-hematite interface. (h) Fe and O relative atomic concentration measured by EELS along the white arrow in panel (g). (i) Monochromated STEM-EELS spectra of Fe-L3,2 fine-edge structure measured on the partially reduced sample in the bulk (cyan) and in the reduced layer closer to the surface (blue curve). The blue arrows mark the three L2 components arising from the three Fe sites in magnetite (octahedral Fe2+ and Fe3+ and tetrahedral Fe3+). The cyan arrow indicates the position of the L3 pre-peak arising from the spin-orbit splitting of the 2p orbital.

**Overview of the previously reported surface reconstructions**

In combination with the bulk configuration, the initial structure and evolution of the surface of hematite determines the performance of a photoelectrocatalytic reaction (e.g., via the amount and type of surface states). Figure 2 shows an overview of individual surface reconstructions typically found on (0001)-oriented $Fe_2O_3$ natural single crystals under different oxygen chemical potentials. They are commonly referred in the literature as (a) (1×1)-$Fe_3O_4$(111), (b) honeycomb/biphase, and (c) (1×1)-$Fe_2O_3$(0001) phases. This nomenclature refers to the magnetite and hematite periodicities relative to the periodicity of the iron layers. Instead, sometimes the periodicities are specified relative to the oxygen basal planes in the literature; namely (2×2)-$Fe_3O_4$(111) and ($\sqrt{3}\times\sqrt{3}$)$R$30°-$Fe_2O_3$(0001). The (1×1)-$Fe_3O_4$(111) phase, Fig. 2a, is obtained when samples are treated under reducing conditions. It results from preferential sputtering of O atoms and subsequent annealing in UHV ($\mu_o$ <−2.5 eV).[1] $Fe_3O_4$(111) may terminate at up to 6 different possible layer cuts, of which the $Fe_{tet}$ bulk termination displayed in Fig. 2a is the most stable as the single phase at $\mu_o$ = −2.5 eV.[43] LEED shows the corresponding diffraction pattern. The Fe 2p XPS spectrum lacks the $Fe^{3+}$ satellite peaks characteristic of $Fe_2O_3$,[32] and shows a distinctive broad $2p_{3/2}$ component due to a mixture of $Fe^{2+}$ and $Fe^{3+}$ multiplet peaks. Fig. 2b shows the honeycomb/biphase reconstruction, which results from annealing in oxygen between $\mu_o$ = −2.5 eV and −1.8 eV. This surface displays complex tip-dependent STM contrasts and a floretted LEED pattern, arising from the Moiré superstructure formed by the $Fe_2O_3$ substrate and an $FeO_2$ overlayer.[24] The associated Fe 2p XPS signal develops $Fe^{3+}$ satellites and a sharper Fe $2p_{3/2}$ component. At higher $\mu_o$, $Fe_2O_3$ natural crystals exhibit instead a (1×1) periodicity, Fig. 2c. The LEED pattern of the (1×1) phase shows only the main spots. The Fe 2p lineshape from the (1×1) phase shows $Fe^{3+}$ satellites typical of stoichiometric hematite.

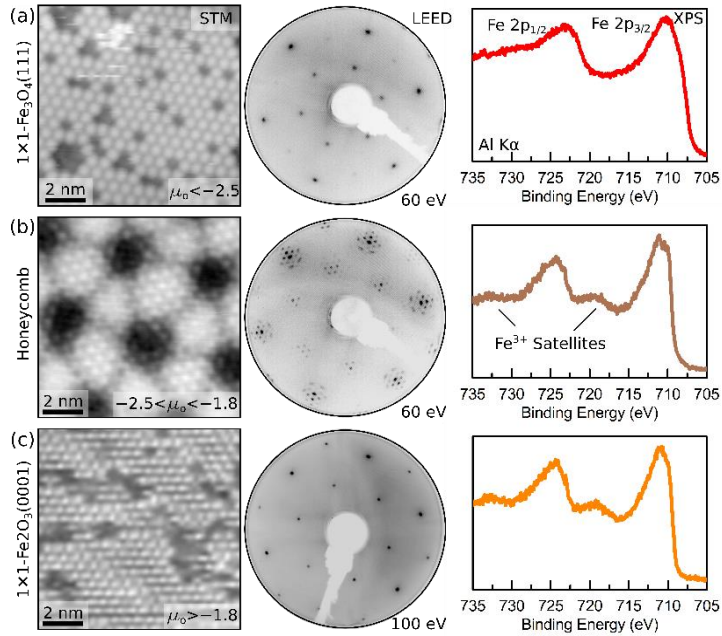

**Figure 2**. Surface reconstructions of α-Fe$_2$O$_3$(0001) natural crystals at different oxygen chemical potentials ($\mu_o$). Representative room-temperature STM, LEED, and XPS results for (a) the (1×1)-Fe$_3$O$_4$(111) reconstruction obtained at $\mu_o < -2.5$ eV; (b) the honeycomb (biphase) reconstruction in the range $-2.5$ eV $< \mu_o < -1.8$ eV. STM and LEED reveal a complex Moiré pattern; (c) the bulk-like termination of Fe$_2$O$_3$ at $\mu_o > -1.8$ eV. STM parameters: (a) -0.5V, 0.5 nA, (b) 1.5 V, 0.6 nA, (c) 1 V, 3 nA. LEED: The pattern in (c) was obtained from a different sample than (a) and (b); the sample was mounted on the sample plate with a different orientation. XPS: $E_{\text{pass}} = 20$ eV

**Is the honeycomb structure a true biphase?**

The controversy around the biphase vs. honeycomb models has its foundation in the complex STM contrast shown in Fig. 2b, which has often been interpreted as a coexistence of multiple distinct structural phases (FeO and Fe$_2$O$_3$). To address this controversy, nc-AFM measurements at 78 K were performed. Figure 3a shows an atomically resolved, constant-height image of the honeycomb phase. Individual atoms, imaged as dark circles, are arranged with a periodicity of $0.30 \pm 0.01$ nm (yellow rhombus). This basic structural motif is modulated on a large scale forming the 4 ± 0.1 nm honeycomb superstructure (green rhombus). Within the superstructure, three areas have a distinct appearance. These are marked as (√3×√3), 'bright' (1×1), and 'dark' (1×1). The fast Fourier transform (FFT) shown in Fig. 3b gives an additional 0.52 ± 0.03 nm periodicity, which is most apparent within the (√3×√3) region (blue rhombus). The unit cell vector with length of 0.52 nm corresponds to the hematite substrate.

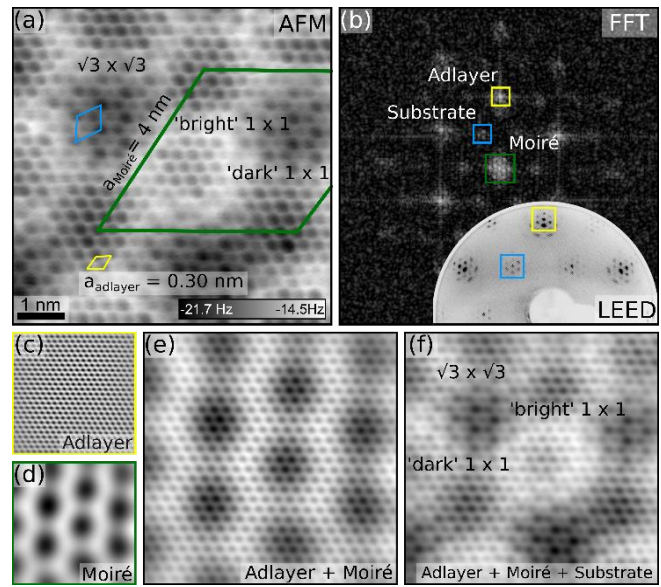

**Figure 3.** Analysis of the honeycomb phase with constant-height nc-AFM. (a) Atomically-resolved image (8 nm × 8 nm). Atoms are imaged in the attractive regime (dark contrast). (b) FFT of the image shown in panel a). The yellow, blue, and green squares highlight the (1×1) (√3×√3) and Moiré periodicities found in the real space, respectively. The inset shows a LEED pattern of the same surface reconstruction. (c) and (d) Adlayer and Moiré components of the nc-AFM image extracted from panel (b) by inverse FFT. (e) Composition made by adding panels (c) and (d), showing an AFM contrast reminiscent of a Moiré pattern. (f) Composition obtained by adding the substrate contribution (blue) to panel (e).

The LEED pattern in Fig. 2b and the FFT in Fig. 3b are strikingly reminiscent of a Moiré structure. It arises from the existence of a large periodicity, as revealed by the nc-AFM contrast in Fig. 3a. Previous work[24] attributed the presence of one (√3×√3) and two (1×1) domains in the STM images to differences in atomic heights of the last O layer of FeO$_2$, according to the overlayer placement on the substrate.[24] However, the different domains could also arise from electronic effects due to different tunneling between tip and sample at different sites.[37] The nc-AFM images support the former scenario. They were obtained in constant height mode (high sensitivity to sub-nanometer atomic-height variations) at the contact potential difference between tip and sample. At these conditions, Pauli and van der Waals interactions (electrostatic contributions) are minimized. Moreover, since hematite is insulating at the acquisition temperature of 78 K, there is no tunneling current between tip and sample. Thus, the measured contrast should largely arise from slight atomic height differences along the honeycomb superstructure, which strongly supports the topographic origin of the AFM (and STM) contrast in favor of the overlayer (honeycomb) model. Figs. 3c and 3d show a decomposition of the nc-AFM image in the adlayer (yellow) and Moiré (green) components of the FFT of Fig. 3b. The combination of Moiré and adlayer contributions in Fig. 3e produces an image typical for supported 2D materials.[44] Adding the substrate contribution to Moiré plus adlayer composition produces the ($\sqrt{3}\times\sqrt{3}$) and two (1×1) fine modulations, given in Fig. 3f. The ($\sqrt{3}\times\sqrt{3}$) and (1×1) areas can be, in principle, identified as 'domains' with distinct local structure and chemistry (the local differences in AFM contrast indicate different tip–sample chemical forces), but not in the extent considered within the biphase model (alternating islands of different stoichiometry, FeO and Fe$_2$O$_3$).

**Forming the honeycomb or (1×1)-Fe$_2$O$_3$(0001): The role of impurities**

The (1×1)-Fe$_2$O$_3$(0001) atomic termination at high $\mu_o$, Fig. 2c, is also subject to some controversy. The [0001] direction of hematite has a polar alternation of Fe and O planes. Based on the autocompensation mechanism, only a stoichiometric Fe-terminated surface would consistent with non-polarity under UHV conditions.[25,45] However, (1×1) O-, Fe-, H-O- and mixed terminations have been reported for samples prepared in UHV and near-ambient pressure (NAP).[1,46] It has been hinted that their stability might be linked to the metallic support used to grow $Fe_2O_3$ films.[27] Hematite single crystals can be considered as a quasi-infinite system on which autocompensation must happen at a vacuum-exposed surface. On supported oxide thin films, this can occur at the metal-oxide interface, providing more degrees of freedom for surface terminations. Another possibility is that the surface structure is altered by natural or incorporated impurities.[34] The (1×1) termination has been reported to form on both natural and synthetic single crystals.[47] However, it cannot be ruled out that the (1×1) forms due to impurities (e.g., adventitious carbon and water) adsorbed during air transport to the measuring instruments. Natural impurities or added dopants such as K and Ti are known to induce surface restructuring on $\alpha$-$Fe_2O_3$($1\bar{1}02$).[32,33]

To disentangle the effects of sample impurities from other effects on given surface reconstructions, natural and synthetic $Fe_2O_3$ sample were prepared and investigated in UHV, using typical ranges of pressure and temperature found in the literature for the preparation of the honeycomb and (1×1)-$Fe_2O_3$(0001) phases. The amount and type of impurities in natural crystals vary with each sample. Common impurities detected during XPS investigations were alkali metals such as Ca, K, and Na, and transition metals such as Ti, Mn and Cr. Clean, epitaxial thin films of ≈100 nm thickness were grown by pulsed laser deposition (PLD) on natural hematite single crystals,[33] henceforth referred to as 'synthetic' hematite. These films are free from impurities within the resolution limit of the XPS setup. Table 1 shows the range of chemical potentials probed in this work when oxygen-annealing natural and synthetic $\alpha$-$Fe_2O_3$(0001) samples. Natural crystals exhibit the honeycomb surface reconstruction when treated under reducing conditions and the (1×1)-$Fe_2O_3$(0001) termination under oxidizing conditions. The phase transition between honeycomb and (1×1)-$Fe_2O_3$(0001) on natural crystals fits well with the $\mu_o$ values reported for these phases on metal-supported hematite and natural crystals. On synthetic samples, however, only the honeycomb reconstruction was obtained; the (1×1)-Fe$_2$O$_3$(0001) termination could not be reproduced within the pressure and temperature ranges that are commonly applied. What is more, the growth kinetics and time used for equilibration of the honeycomb phase depends on $\mu_o$. To obtain the honeycomb phase from the Fe$_3$O$_4$-terminated hematite within UHV-compatible oxygen pressures ($p \lessapprox 10^{-4}$ mbar), annealing between 2 and 8 hours is typically required (the exact duration depends on the history of the sample). However, merely 10−30 min are required under high oxygen pressure (≈ 1 mbar).

| $\mu_o$ (eV) | Natural | Synthetic | $p$(mbar) | T (°C) |
|---|---|---|---|---|
| −2.22 | Honeycomb | Honeycomb | 1×10$^{-9}$ | 700 |
| −2.04 | Honeycomb | Honeycomb | 1×10$^{-5}$ | 800 |
| −1.94 | Honeycomb | Honeycomb | 1×10$^{-6}$ | 700 |
| −1.84 | Honeycomb | Honeycomb | 1×10$^{-5}$ | 700 |
| −1.74 | (1×1) α-Fe$_2$O$_3$(0001) | Honeycomb | 1×10$^{-4}$ | 700 |
| −1.63 | (1×1) α-Fe$_2$O$_3$(0001) | Honeycomb | 1×10$^{-5}$ | 600 |
| −0.92 | (1×1) α-Fe$_2$O$_3$(0001) | Honeycomb | 1×10$^{-4}$ | 300 |
| −0.77 | (1×1) α-Fe$_2$O$_3$(0001) | Honeycomb | 0.5 | 300 |
| −0.55 | (1×1) α-Fe$_2$O$_3$(0001) | Honeycomb | 0.5 | 150 |

**Table 1.** Summary of surface structures obtained when equilibrating natural and synthetic Fe$_2$O$_3$ at different oxygen chemical potentials. The honeycomb-to-(1×1) transition occurs at a $\mu_o \sim -1.8$ eV only on natural samples; however, the precise value depends on the specific sample.

**Contamination segregation during hematite oxidation.**

The effect of the intrinsic impurities of natural crystals not only affect the stability of surface reconstructions of α-Fe$_2$O$_3$(0001), but also the bulk transformations and the fully stoichiometric recovery of hematite. Fig. S1a in the Supplementary Information shows a HRTEM image of a partially reoxidized magnetite-like inclusion within the fully recovered hematite. FFT analysis of the inclusion and the surrounding area, Figs. S1b and c, reveals that the inclusion has the same magnetite/maghemite structure as shown in Fig. 3c, whereas the rest of the bulk has the characteristic hematite structure. The EELS mapping in Figs. S1d and S1e reveals an inhomogeneous presence of Mn in the 6 - 12% range. The Fe concentration is locally decreased in an equivalent amount. O concentration is decreased only to a degree expected for magnetite. The Fe-$L_{2,3}$ edge structure of the inclusion, Fig. S2, resembles that of magnetite reported in Fig. 1i. This evidence suggests the formation of $Fe_{3-x}Mn_xO_4$ ferrite. Outside the inclusion, the Mn concentration is below the detection limit of EELS. The interplanar-spacing vector lengths of the Mn-rich area is shown in Table ST1 in the Supplementary Information. Mn locally hinders the fully reoxidation of the sample at oxygen chemical potentials at which a fully stoichiometric hematite is obtained on clean samples. Moreover, Mn single dopants alter the hole mobility in hematite[48] and, depending on the Mn concentration and sample preparation parameters, Mn possibly could lead to ferrite or Mn oxide formation. It is also likely that minute amounts of impurities such as Mn can distribute all over the hematite surface, resulting in a (1×1) bulk-like termination.

These results clearly show that the natural impurities play a crucial role in changing the thermodynamic balance of the hematite surface and near-surface structure and chemistry. During the reoxidation, the impurities are pushed out, thus changing the surface and near-surface composition and chemistry. The concentration and elemental distribution of contaminants vary within each natural crystal and preparation. Hence, synthetic films are essential for an adequate characterization of hematite in catalysis.

**Conclusion**

This work addressed the structure of the hematite $\alpha$-$Fe_2O_3$(0001) surface and near-surface. High-resolution nc-AFM images acquired in UHV strongly support that the oxidized surface of hematite is natively composed of a 2D oxide layer. The layer shows a Moiré pattern with three different areas due to the distinct attachment to the underlying fully oxidized bulk; these areas are responsible for the 'biphase' contrast observed in STM. A series of controlled experiments compared the oxidation of a synthetic films with natural crystals, the former clean within the resolution limit of XPS, and the latter contaminated by intrinsic impurities. The surface termination is strongly influenced by the presence of impurities, forcing the formation of surface phases that are not thermodynamically preferred on a clean sample. Specifically, the presence of impurities (e.g., alkali metals) enforces the formation of a (1×1)-$Fe_2O_3$(0001) termination instead of the honeycomb one in a wide range of oxygen chemical potentials. Finally, subsurface cross-sectional imaging by TEM reveals the spatial extent of changes introduced by surface preparation methods and the spatial localization of impurities that can be associated only with a particular ferrite phase.

The atomic structure and transformations of $\alpha$-$Fe_2O_3$(0001) during actual catalytic reactions are generally poorly understood.[36] Impurities significantly influence the structure and chemistry of both the hematite surface and its bulk. This is of relevance in photoelectrocatalysis (water splitting, Fenton process, Fischer-Tropsch reaction), which typically uses alkali-containing electrolytes (e.g., KOH, NaOH, and carbonates) and metal-doped hematite (e.g., Ti, Zn, Mn, Ni) to improve conductivity and charge separation. This study suggests that the surface and bulk of polar hematite facets will be stabilized by the photoelectrochemical conditions away from the initial state. The fact that the surface structure depends on the environment (both in terms of oxygen chemical potential and presence of alkali metal atoms) provides the essential clue to interpret the experimental results on the catalytic activity of hematite surfaces. Metallic dopants, surface (oxy)hydroxylation and/or cation adsorption will play a crucial role in achieving a non-polar, stable surface structure of $\alpha$-$Fe_2O_3$(0001), i.e., the actual catalytically active phase. This should be considered when aiming to rationalize structure-function relationships.

**Methods**

**Sample preparation:**

Natural $\alpha$-Fe$_2$O$_3$(0001) single crystals were acquired from SurfaceNet GmbH. A total of ten crystals have been used to carry out the experiments. Each crystal had a different concentration and elemental composition of impurities; the ones found were: Na, Ca, K, Mn, Zn, In, Ti, Mo, Cr, Al, Sr.

To reduce the samples into the magnetite phase, $\alpha$-Fe$_2$O$_3$(0001) single crystals were sputtered with Ar$^+$ (10 min, 1 × 10$^{-6}$ mbar, 1−1.5 keV, 10 mA, 60 µA) and annealed in UHV (600−800 °C, 30 min) until a clear (1×1) magnetite LEED diffraction pattern was observed with no hematite diffraction spots. The honeycomb phase was obtained by oxygen annealing (1 × 10$^{-6}$ mbar, 500−800 °C, 1−8h) until no magnetite phase was detected by LEED. The (1×1) phase was obtained on natural crystals after producing the honeycomb by oxygen annealing (1 × 10$^{-5}$ mbar of O$_2$, 500°C, 30 min). The annealing was performed by radiative heating from a hot filament (Ta) and a temperature increase of 50−100 °C/min to avoid sample cracking. During the measurements, the samples were regularly refreshed by a 10−20 min oxygen annealing (1 × 10$^{-6}$ mbar, 600 °C) to remove adventitious carbon contamination. Two crystals were heavily contaminated by bulk dopants, and it was not possible to reoxidize the magnetite phase.

**Synthetic film growth:**

Synthetic hematite was obtained by PLD in a dedicated setup with in-UHV transfer to a dedicated surface-science chamber[49]. Films of roughly 100 nm thickness were grown on natural $\alpha$-Fe$_2$O$_3$(0001) single crystals from a single-crystalline Fe$_3$O$_4$ target (700 °C, 2 × 10$^{-2}$ mbar O$_2$, 5 Hz, 2.0 J/cm$^2$, no post-annealing, 60 °C/min ramp rate), as detailed elsewhere[50]. Before the growth, the natural samples were cleaned by repeated sputtering-annealing cycles (10 min, $1 \times 10^{-6}$ mbar Ar$^+$, 1 keV; 30min, 600 °C, $1 \times 10^{-6}$ mbar O$_2$) till no change in the contaminant signals was visible in XPS. Then, the samples were annealed at 850 °C for 1 h at $2 \times 10^{-2}$ mbar O$_2$, to promote the flattening of the surface morphology and ensure complete oxidation of the crystals. The cleanliness of the films was checked by LEED and XPS.

**XPS, LEED, STM and nc-AFM characterization:**

Normal emission XPS measurements were carried out using a laboratory-based system (SPECS Surface Nano Analysis GmbH, monochromatized Al-K$\alpha$ source) with a base pressure of $5 \times 10^{-10}$ mbar. The core-level spectra were recorded with a pass energy of 20 eV, step size of 0.05 eV, and dwell time of 200 ms. No charging was observed. LEED patterns were acquired with SPECS ErLEED 150 setups. The LEED apparatus was under operation parameters for at least 1 h before transferring the samples to avoid contamination (C, F). High-energy (150−300 eV) electrons were used to probe traces of the magnetite phase below the surface layers. Room-temperature STM images were acquired in several STM apparatuses with electrochemically etched W tips. The tips were treated on Au(111) before measuring hematite. Non-contact AFM images were acquired in an Omicron POLAR-SPM microscope at 78 K using Qplus® sensors (resonance frequency of ca. 47 kHz, $Q$ factor of ca. 5000) with an electrochemically etched W tip. The tip was prepared on Cu(100) until a change in the resonance frequency smaller than $-1.5$ Hz at 0.1 V bias was obtained while approaching the tip, and the contact potential difference between tip and sample was < 0.2 V.

**Data processing:**

XPS data was processed using the KolXPD software. The binding energy positions were calibrated using the fermi edge measured on a Ta sample plate. LEED images were acquired by averaging for 10 s with a camera in a dark receptacle. Dark-field images were acquired by turning off the LEED-screen acceleration voltage. They were subtracted from the original data to remove stray light and filament reflections. The contrast of the images was inverted to enhance the diffraction pattern. The STM and nc-AFM images were processed using custom ImageJ plug-ins. Microscope noise frequencies were filtered out. The lattice parameters of the honeycomb phase were obtained by dividing the distance between two spots containing 10-15 atomic positions by the number of unit cells. This was repeated in several locations in the three crystallographic directions to account for scanning distortions. The error bars correspond to the standard deviation of these measurements. The figures were prepared using ImageJ, Gimp and Inkscape.

**TEM lamella fabrication and measurements:**

The samples were covered by an Ag layer in the same UHV system were the samples were prepared. This protective cap prevents contamination of the topmost surface layers as a result of exposure to ambient conditions. Additionally, it prevents the surface from coming in direct contact with the reactive layers that are deposited during the lamella fabrication. Ag was deposited by thermal evaporation from a Knudsen-type effusion cell (crucible at 850 °C). The sample was held at room temperature in front of the effusion cell for 78 min, resulting in a ~30 nm-thick Ag film. The pressure in the chamber during deposition was $4 \times 10^{-10}$ mbar. Afterwards, the sample was removed from UHV.

As the first step of the lamella fabrication, a protective cap (~ 350 nm thick) was deposited using electron-beam-induced deposition (EBID), followed by ion-beam-induced deposition (IBID) of the same element. Several gas injection system (GIS) chemistries for the deposition were tested (C, Pt, W): W or Pt did not introduce unwanted species into the layers of interest. The deposition of C makes it difficult to restore the sample for UHV experiments and causes extensive deposition of carbonaceous layers during TEM/EELS measurements, which are detrimental to the quality of results. The lamella was then liberated from the bulk by FIB milling with 30 keV Ga ions and transferred onto a Cu support grid. In order to suppress carbon contamination of the lamella as much as possible, the FIB-SEM chamber was plasma-cleaned before inserting the bulk sample. Two final thinning and polishing steps were conducted at 5 and 2 keV beam energies, respectively. In this study, the TEM lamella surface is normal w.r.t. the $[11\bar{2}0]$ direction of bulk hematite.

The TEM measurements were performed at an accelerating voltage of 300 kV with a microscope TITAN Themis 60-300 (Thermofisher Scientific) equipped with a monochromator and a spherical aberration ($C_S$) corrector of objective lens. The HRTEM images were acquired with $C_S \sim 0$ µm and with an appropriate defocus in range of few nm to observe atomic columns with minimum delocalization. Velox software v.2.12 was used for the image acquisition and processing of corresponding FFT patterns used for crystallographic evaluation.

The STEM-EELS measurements were performed with a Quantum ERS spectrometer (Gatan), a probe convergence semi-angle of 10 mrad, collection semi-angle of 28.2 or 56.4 mrad and entrance aperture of 2.5 or 5 mm, respectively. The EELS datasets were obtained with GMS v.3.3 software with enabled Dual-EELS mode allowing a simultaneous collection of a low-loss spectrum image and a high-loss spectrum image containing a zero-loss peak and edges of elements of interest, respectively, in each pixel. The used pixel size (i.e., the spatial resolution) and pixel time of the high-loss spectrum images was 0.3 – 0.5 nm and 0.02 – 0.08 s, respectively. The EELS data of Fe-$L_{3,2}$ fine-edge structure was acquired in monochromated STEM and with electron energy dispersion of 0.025 eV per channel giving an energy resolution approximately 0.12 eV. The EELS data for chemical concentration measurements and elemental mapping were performed with dispersion of 0.25 or 0.5 eV per channel. Relative chemical concentrations were calculated in At. % by using a model-based EELS quantification function included in the GMS 3, where the following settings were chosen with emphasis for the best fit with the obtained spectra: signal sum width was selected within extended energy-loss fine structure part of particular edges, no overlap of edges was selected, background subtraction was performed with a Power Law model (for Fe-L and O-K edge) and a 1st Order Log-Polynomial model (for Mn-L edge), cross-section was calculated with a Hartree-Slater model and plural scattering was deconvoluted from high-loss spectrum images with the use of corresponding low-loss spectrum images. The Fe-$L_{3,2}$ fine-edge structure analysis was performed after the background subtraction by the Power Law model and plural scattering deconvolution by a Fourier-Ratio method.

ASSOCIATED CONTENT

**Supporting Information**.

Mn-rich inclusion structural and chemical identification.

AUTHOR INFORMATION

Jesús Redondo, redondo@karlov.mff.cuni.cz, ORCID:0000-0002-8147-689X, Tw:@JesusRedox

Jan Michalička, jan.michalicka@ceitec.vutbr.cz, ORCID: 0000-0001-6231-0061

Jan Čechal, cechal@fme.vutbr.cz, ORCID:0000-0003-4745-8441

Giada Franceschi, franceschi@iap.tuwien.ac.at, ORCID:0000-0003-3525-5399

Gareth Parkinson, parkinson@iap.tuwien.ac.at, ORCID:0000-0003-2457-8977

Martin Švec, svec@fzu.cz, ORCID:0000-0003-0369-8144

Nishant Kumar, nishant.kumar@ceitec.vutbr.cz, ORCID:0000-0002-1021-2635

Matthias Blatnik, matthias.blatnik@ceitec.vutbr.cz, ORCID: 0000-0001-5448-8580

Florian Kraushofer, florian.kraushofer@tum.de, ORCID:0000-0003-1314-9149

Michele Riva, riva@iap.tuwien.ac.at, ORCID:0000-0001-8303-7383

Ulrike Diebold, diebold@iap.tuwien.ac.at, ORCID:0000-0003-0319-5256

Notes: The authors declare no competing financial interest.

ACKNOWLEDGEMENT

J.R. acknowledges support from the Grant Agency of Charles University (GAUK) and Czech Republic (GAČR) projects number 160119 and 22-18079O, respectively. G.F. and U.D. were supported by the European Research Council (ERC) under the European Union's Horizon 2020 research and innovation programme, grant agreement No. 883395, Advanced Research Grant 'WatFun'. G.S.P. and F.K. were supported by the ERC under the European Union's Horizon 2020 research and innovation programme, grant agreement No. 86462, Consolidator Research Grant 'E-SAC'. Part of the research was carried out using the CzechNanoLab Research Infrastructure supported by MEYS CR (LM2018110). J.Č. was supported by GAČR, project No. 22-05114S. N.K. acknowledges Grant CEITEC VUT-K-22-7782, realized within KInG BUT scheme, No. CZ. 02.2.69/0.0/0.0/19_073/0016948, financed from the OP RDE. M.B. acknowledges financial support through the ERC and MEYS CR co-founded IMPROVE V project CZ.02.01.01/00/22_010/0002552. M.S. acknowledges the support by GAČR, project No. 22-18718S. M.R. acknowledges support from the Austrian Science Fund (FWF) through project SFB-F81 "Taming Complexity in materials modeling" (TACO). The authors thank Michael Schmid for providing ImageJ plug-ins.

# Supplementary Information – Hematite α-Fe$_2$O$_3$(0001) in top and side view: resolving long-standing controversies about its surface structure

*Jesús Redondo[1,2]\*, Jan Michalička[3], Giada Franceschi[4], Břetislav Šmid[1], Nishant Kumar[3], Ondřej Man[3], Matthias Blatnik[3], Dominik Wrana[1], Florian Kraushofer[4], Benjamin Mallada[5], Martin Švec[5], Gareth S. Parkinson[4], Martin Setvín[1], Michele Riva[4], Ulrike Diebold[4]\*, Jan Čechal[3]\**

AUTHOR ADDRESS:

1 Charles University, Faculty of Mathematics and Physics, Prague, Czech Republic

2 University of the Basque Country, Faculty of Chemistry, Donostia-San Sebastián, Spain

3 Central European Institute of Technology, Brno, Czech Republic

4 Institute for Applied Physics, Technical University Wien, Vienna, Austria

5 Institute of Physics, Czech academy of Science, Prague, Czech Republic

AUTHOR INFORMATION

**Corresponding Author**

\*Jesús Redondo redondo@karlov.mff.cuni.cz.

\*Ulrike Diebold diebold@iap.tuwien.ac.at.

\*Jan Čechal cechal@fme.vutbr.cz

# Mn-rich clusters on natural α-Fe$_2$O$_3$(0001)

Intrinsic impurities on natural Fe$_2$O$_3$ crystals can be pushed to the surface and near-surface regions at the temperatures and oxygen partial pressures that are used to obtain stoichiometric hematite under UHV. Figure S1 shows a representative case where Mn impurities form clusters with stoichiometry and structure close to inverse spinel Fe$_3$O$_4$. These clusters are embedded within stoichiometric hematite and are not oxidized to Fe$_2$O$_3$ within the temperature and pressure range achievable under UHV.

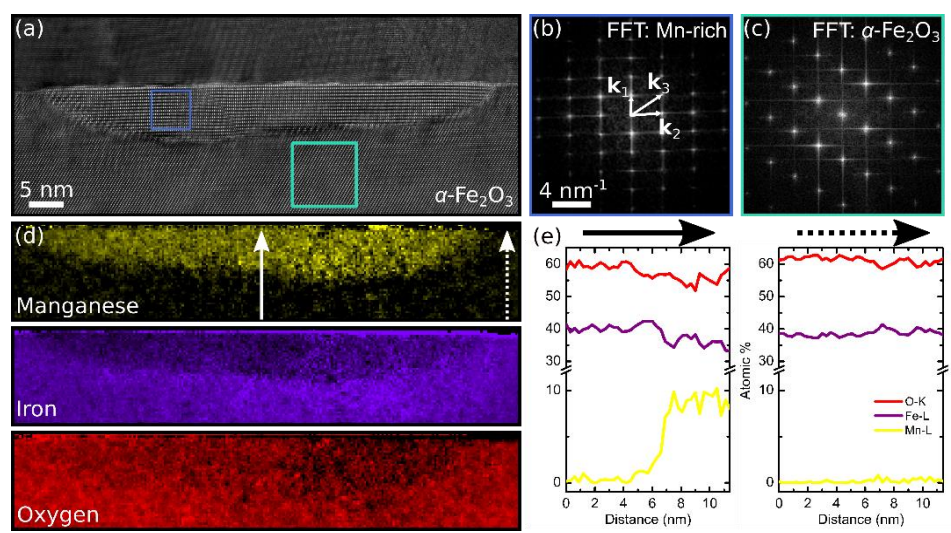

**Figure S1.** (a) HRTEM image of a Mn-rich cluster within hematite. (b) and (c) FFT analysis of the inclusion (blue) and the bulk (cyan) revealing magnetite- and hematite-like symmetries, respectively. (d) STEM-EELS chemical mapping of the inclusion showing the distribution of Mn, Fe, and O. (e) Relative atomic concentration of Mn, Fe, and O obtained from the intensity of the Mn-L$_{3,2}$, Fe-L$_{3,2}$, and O-K lines measured by STEM-EELS along the white dotted and solid arrows in (a).

The interplanar-spacing vector lengths $d$ (Table ST1) obtained from the FFT of the Mn-rich phase (Fig. S1b) reveal a unit cell close within the experimental error to the structure of magnetite observed in Fig. 3b of the main text and reported elsewhere.[1] The relative atomic concentrations shown in Fig S1e indicate a 40:60 Fe to O ratio in the Mn-free areas corresponding to hematite. The Mn-rich clusters shows a 6-12% Mn concentration.

| FFT | d (measured) [Å] | d (magnetite in Fig. 3b) [Å] | $d_{hkl}$ ($Fe_3O_4$ ref.) [Å] | hkl ($Fe_3O_4$) |
|---|---|---|---|---|
| $k_1$ | 4.95 | 4.90 | 4.84 | 111 |
| $k_2$ | 3.03 | 2.98 | 2.97 | 022 |
| $k_3$ | 2.56 | 2.57 | 2.53 | 113 |

**Table ST1.** *d*-spacings found for the Mn-rich inclusion compared to *d*-spacings of the magnetite phase in Figure 3b of the main manuscript and the literature.[1]

The Fe-$L_{3,2}$ fine-edge structures in Fig. S2 measured in bulk (dark curve) and the Mn-rich cluster (red curve) are reminiscent of the hematite and magnetite spectra shown in Fig. 3i in the main manuscript, respectively. The split $L_3$ pre-peak measured from the bulk region (marked by a black arrow) allows the identification of a hematite phase, whereas the $L_2$ peak (red arrows) measured in the Mn-rich area are characteristic of magnetite.[2] The presence of Mn and the structure and stoichiometry resembling that of magnetite (cubic inverse spinel) hints towards the formation of a ferrite $Mn_xFe_{2-x}O_3$ phase embedded in the hematite crystal surface.

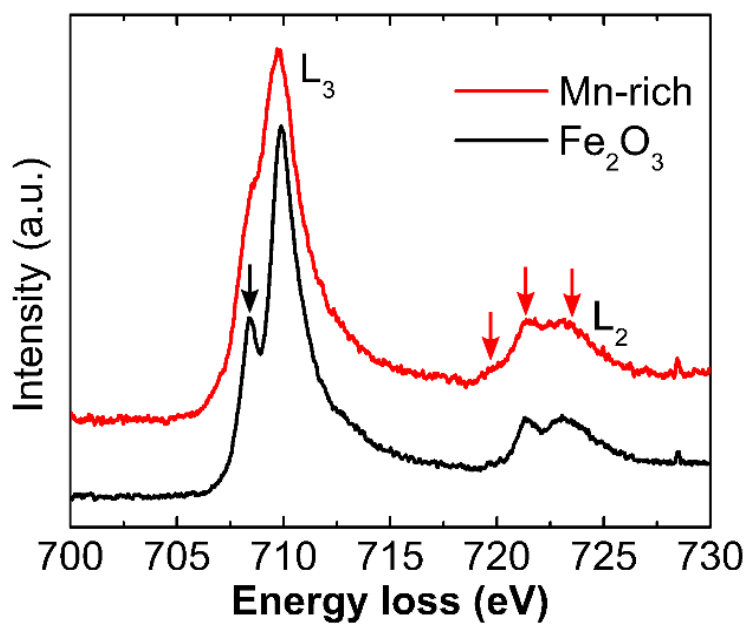

**Figure S2.** STEM-EELS of a Mn-rich region (green) and bulk (red) showing the Fe $L_3$ and $L_2$ edge fine structure of magnetite and hematite, respectively.